\documentclass[pre,twocolumn,showpacs]{revtex4}

\usepackage{graphicx}
\usepackage{bm}

\newcommand{\avec}[1]{{\bm{#1}}}
\newcommand{\avecu}[1]{{\hat{\bm{#1}}}}

\begin{document}

\title{Radial and axial segregation of granular matter in a rotating cylinder: A
simulation study}

\author{D. C. Rapaport}
\email{rapaport@mail.biu.ac.il}
\affiliation{Physics Department, Bar-Ilan University, Ramat-Gan 52900, Israel}

\date{December 26, 2006}

\begin{abstract}

The phenomena of radial and axial segregation in a horizontal rotating cylinder
containing a mixture of granular particles of two different species have been
modeled using discrete particle simulation. Space-time plots and detailed
imagery provide a comprehensive description of what occurs in the system. As is
the case experimentally, the nature of the segregation depends on the parameters
defining the problem; the radial component of the segregation may be transient
or long-lasting, and the axial component may or may not develop. Simulations
displaying the different kinds of behavior are described and the particle
dynamics associated with the axially segregated state examined. The importance
of an appropriate choice of interaction for representing the effective friction
force is demonstrated.

\end{abstract}

\pacs{45.70.Mg, 45.70.Qj, 64.75.+g, 02.70.Ns}

\maketitle

\section{Introduction}

The mechanisms responsible for some of the more surprising features of granular
matter continue to challenge understanding. In view of the fact that the
behavior encountered often differs strongly from systems governed by statistical
mechanics and thermodynamics, there is little by way of intuitive help from such
theories in trying to understand granular matter, both at rest and in motion.
Segregation is perhaps the most prominent of these features; the ability of
noncohesive granular mixtures to segregate into individual species, despite the
absence of any obvious entropic or energetic benefit, makes this a particularly
fascinating phenomenon.

Granular separation and mixing are central to many kinds of industrial
processing that span a broad range of size scales; the capability for either
causing or preventing segregation can be central to the viability of a
particular process. Substantial economic benefits could be gained from a
systematic understanding of the complexities of granular flow at a level similar
to fluid dynamics; the outcome would be improved control, reliability and
efficiency, a notable advance over currently used approaches that are often
empirical. Analogous segregation processes occur in nature, where an
understanding of the underlying mechanisms might benefit resource management.
Consequently, considerable effort has been invested in exploring these
phenomena, but, absent a general theory of granular matter, much of the progress
in this field \cite{bar94,her95,jae96,kad99} relies on computer simulation.

There are a variety of conditions under which granular mixtures segregate; these
include shear \cite{wal83,hir97}, vibration \cite{ros87,gal96,rap01} and
rotation \cite{can95,zik94,rap02}. Experiments on rotational segregation, one of
the more extensively studied systems experimentally and the subject of the
present paper, employ a horizontal cylinder revolving at constant rate and
partially filled with a mixture of two species of granular particles. Given the
appropriate kinds of particles, and under suitable conditions, the mixture
spontaneously segregates into axial bands of alternating composition. A
substantial body of experimental data has been assembled, but because of the
numerous parameters entering the experiment only a limited portion of the
multidimensional phase diagram has been explored. The relevant parameters
include the nature of the granular particles themselves, ranging from the most
familiar forms of granular matter such as sand, through various organic
products, plastic and glass beads, to metal ball bearings. The particles have a
broad spectrum of elastic and frictional characteristics, and sizes that range
over several orders of magnitude. The material can be in a dry or wet state, the
latter case known as a slurry. The cylinder can have different shapes, most
commonly one with a uniform circular cross section. Further parameters include
the cylinder diameter and length, the rotation rate, the relative fractions and
sizes of the granular components, and the filling level of the container. In
marked distinction to fluid systems, there are no known scaling relations that
allow this parameter space to be reduced, and while some systematic trends have
been noted, there is little guidance available for predicting the outcome of an
experiment involving some unexplored combination of parameters.

The most direct observations address the structure of the upper surface
\cite{zik94}, while more elaborate studies examine the interior using, for
example, noninvasive magnetic resonance imaging (MRI)
\cite{hil97s,hil97f,nak97}. Axial segregation may be complete, with each band
composed exclusively of one particle species, or it may be only partial, with
bands characterized by higher concentrations of alternating species. There is an
additional effect that occurs entirely beneath the surface leaving little
external evidence, namely radial segregation, which produces a core rich in
small particles extending along the length of the cylinder. Radial segregation
occurs early in experiments starting from a uniformly mixed state, and may or
may not persist once axial segregation appears. Axial band formation might be
regarded as a consequence of this core intermittently bulging by an amount
sufficient to reach the outer surface \cite{hil97f}. The axial band patterns can
also exhibit time dependence; examples of such behavior involve coarsening, in
which narrower bands gradually merge into broader bands \cite{fre97}, and the
appearance of traveling wave patterns on the surface \cite{cho97,cho98}. Further
discussion of the experimental history appears in \cite{hil97f,sta98,rap02}, but
information about this phenomenon still continues to accumulate. More recent
results showing the richness of the segregation effect and its dynamics in dry
and/or wet mixtures (slurries exhibit behavior that is similar in many respects,
despite the lubricating role of the fluid medium) are described in
\cite{fie03,arn05,fin06}. The problem can be extended in different directions,
for example \cite{new04} considers mixtures with three and more granular
components. That there is little opportunity for scaling to play a role in
understanding this phenomenon is demonstrated in \cite{ale04}, where the ratio
of the cylinder and particle diameters determines whether axial segregation
occurs and if its appearance depends reversibly on rotation rate; the fact that
the wavelength of the axial pattern does not scale with cylinder diameter is
described in \cite{cha06}.

Given the complexity of the experimental situation and the many lacunae in
parameter space, it is hardly surprising that theoretical progress has been
limited. Continuous one-dimensional models have been developed \cite{ara99} in
which the dynamical variables are the local concentration difference and the
slope of the free surface; these describe the early stages of segregation with
traveling bands and subsequent band merging, although experiment \cite{kha04}
has questioned their ability to characterize the process correctly, and the
obviously important three-dimensional aspects of the phenomenon are absent. The
problem has also been studied using a cellular automata approach based on a
highly simplified model \cite{yan99}.

The most detailed theoretical approach involves direct modeling of the dynamics
of granular particles, employing the same computational methods used in
molecular dynamics simulation of atomistic systems.
While there have been a considerable number of such studies, covering a variety
of granular phenomena, there have been very few simulations of granular
particles in a revolving cylinder exhibiting axial segregation. The first of
these was an early, rather limited treatment \cite{sho98}. This was followed by a
broader study \cite{rap02}, that is extended by the present work, in which the
formation of multiple axial bands was demonstrated, and various aspects of the
segregation phenomenon, such as band merging, the absence of mixing in
presegregated systems, and the manner in which the behavior depended on the
choice of parameters, were investigated. The model differed from that of the
present paper in that the transverse restoring force (discussed in detail below)
was absent and a Lennard-Jones type of normal repulsion was used rather than a
linear force. Exactly the same kind of axial band formation was subsequently
reported in \cite{tab04} (although this similarity was not explicitly noted)
using the same kind of model, but with the linear force.
The appearance of radial segregation is described in
\cite{dur97}, but only for a two-dimensional system.

The purpose of the present paper is to analyze a discrete-particle model that
exhibits both axial and radial segregation, in which the appearance of the
radial effect precedes the axial, and then either disappears or persists as an
axial core. Experimentally, it is the smaller particles that are found to
congregate in the interior, and this feature is reproduced by the simulations;
one issue that was not resolved in the earlier study \cite{rap02} was the fact
that the opposite behavior was encountered, but it will be shown that this is
due to the choice of interparticle forces. The goal here is not a systematic
coverage of parameter space, since the computational resources for this would be
considerable, but rather a demonstration of the typical kinds of outcome with
just a few examples of parameter dependence; indeed, due to a lack of detailed
reproducibility under different initial conditions, as will become apparent
subsequently, obtaining a comprehensive picture of the ``average'' behavior
would require multiple simulation runs for each parameter set. Typically, the
majority of runs described here involve larger systems than before, with a
bigger particle size difference and a more slowly rotating cylinder; these
changes, together with the additional work required by the force computations,
called for a greater computational effort than previously.

\section{Simulation methodology}

\subsection{Granular models}

The models used for granular simulation are generally based on spherical
particles whose resistance to overlap is expressed in terms of a continuous
potential function \cite{cun79,wal83,haf86}. In addition to this excluded-volume
repulsion, particles are subject to forces that aim to replicate the effects of
inelasticity and friction. There are limitations to the accuracy of such models,
with one empirical measure of success being the degree to which the essence of
the phenomenon under study is captured.

Consider a pair of granular particles $i$ and $j$ with diameters $d_i$ and
$d_j$, respectively. The repulsive force between particles depends linearly on
their overlap,
\begin{equation}
\avec{f}_v = k_n \left( d_{ij} - r_{ij} \right) \avecu{r}_{ij} ,
\qquad r_{ij} < d_{ij}, \label{eq:fv}
\end{equation}
where $\avec{r}_{ij} = \avec{r}_i - \avec{r}_j$ is the particle separation,
$r_{ij} = |\avec{r}_{ij}|$, and $d_{ij} = (d_i + d_j) / 2$ is the effective
diameter. An alternative is the Hertz interaction that depends on the 3/2-power
of the overlap \cite{sch96}, whereas in \cite{rap02} the repulsive part of the
Lennard-Jones interaction was used. Dissipative forces act for the duration of
each collision. The first of these is velocity-dependent damping along the line
between particle centers,
\begin{equation}
\avec{f}_d = - \gamma_n (\avecu{r}_{ij} \cdot \avec{v}_{ij}) \avecu{r}_{ij} ,
\label{eq:fd}
\end{equation}
that depends on the component of relative velocity $\avec{v}_{ij} = \avec{v}_i -
\avec{v}_j$ parallel to $\avec{r}_{ij}$. Here, $\gamma_n$ is the normal damping
coefficient, assumed to be the same for all particles. The total normal force
between particles is then $\avec{f}_n = \avec{f}_v + \avec{f}_d$.

Frictional damping acts transversely at the point of contact to oppose sliding
while particles are within interaction range,
\begin{equation}
\avec{f}_s = - \min \left( \gamma_s^{c_i c_j} |\avec{v}_{ij}^s|, \,
\mu^{c_i c_j} |\avec{f}_n| \right) \avecu{v}_{ij}^s , \label{eq:fs}
\end{equation}
where the relative transverse velocity of the particle surfaces at this point,
allowing for particle rotation, is
\begin{equation}
\avec{v}_{ij}^s = \avec{v}_{ij} - (\avecu{r}_{ij} \cdot \avec{v}_{ij})
\avecu{r}_{ij} - \left( {d_i \avec{\omega}_i + d_j \avec{\omega}_j \over
d_i + d_j} \right) \times \avec{r}_{ij} ,
\end{equation}
and $\avec{\omega}_i$ is the particle angular velocity. The value of
$\gamma_s^{c_i c_j}$ depends on the particle types $c_i$ and $c_j$, and
$\mu^{c_i c_j}$ is the static friction coefficient that limits the transverse
force to a value dependent on $|\avec{f}_n|$.

In a model of this type there is no true static friction, a practical
consequence of which would be stick-slip motion. A way of at least partially
overcoming this limitation, though not strictly a correct means of incorporating
the effect, is to introduce a tangential restoring force that acts during the
collision and depends on the cumulative relative displacement at the point of
contact \cite{cun79,wal83,dur97}.
This force has the form $\avec{f}_g = - k_g
\avec{u}_{ij}$, where
\begin{equation}
\avec{u}_{ij} = \int_{\rm (coll)} \avec{v}_{ij}^s (\tau) \, d \tau
\label{eq:udef}
\end{equation}
is evaluated as a sum of vector displacements over the interval the particles
have been in contact; the magnitude of $\avec{f}_g$ is also limited by $\mu^{c_i
c_j} |\avec{f}_n|$. In addition, to avoid occasional unrealistically large
displacements (although this does not appear to affect the behavior), if
$|\avec{u}_{ij}| > 0.1$, an arbitrarily chosen limit, it is reset to zero (a
form of ratcheting intuitively motivated by the asperities responsible for
friction); an alternative treatment appears in \cite{sil01}. This
history-dependent force was not present in \cite{rap02}; the results below
suggest that it plays an important role in achieving the correct form of radial
segregation. The total transverse force is $\avec{f}_t = \avec{f}_s + \avec{f}_g$.

The curved cylinder wall and the flat end caps are treated as rough and smooth
boundaries respectively. The interaction of particles with the curved boundary
(together with gravity) drives the system, so this force includes the same
components as the interparticle force. On the other hand, to minimize spurious
effects associated with the end caps, only $\avec{f}_v$ and $\avec{f}_d$ act
there. Further details concerning the interactions (with the exception of
$\avec{f}_g$) appear in \cite{rap02}, together with a discussion of the friction
coefficients and the particle-wall force computations. Other aspects of the
simulation follow standard molecular dynamics procedures \cite{rap04}; neighbor
lists are used to efficiently organize the force computations, the translational
and rotational equations of motion are integrated with the leapfrog method, and
parallel processing can improve performance.

\subsection{Parameters}

Those parameters that also appeared in \cite{rap02} have been assigned similar
values here. The first of these is the gravitational acceleration, $g = 5$,
which then relates the dimensionless MD units used in the simulation to the
corresponding physical units. Thus, if $L_{MD}$ is the length unit (in mm), then
the time unit is $T_{MD} \approx 10^{-2} \sqrt{5 L_{MD}}$\,s. If the cylinder
rotates with angular velocity $\Omega$ (MD units) the actual rotation rate is
$\Omega / (2 \pi T_{MD}) \approx 7.1 \Omega / \sqrt{L_{MD}}$\,Hz; for 3\,mm
particles, $\Omega = 0.1$ is equivalent to 25\,rpm, a typical experimental
value.

Axial segregation tends to develop slowly, leading to long simulation runs.
Computational cost can be reduced by using the largest possible integration time
step; the step size $\delta t$ is limited by the highest particle speeds
encountered and here the value used is $\delta t = 5 \times 10^{-3}$ (MD units).
Numerical stability considerations then set a lower bound on the collision time
to ensure it remains much larger than $\delta t$ (typically 30--40$\times$).
Collisions between particles in the slowly varying bulk interior can be
protracted events, but the opposite is true for particles bouncing rapidly along
the upper free surface, and this in turn constrains coefficients such as
particle stiffness. For the present work, most runs use $k_n = 1000$ in
Eq.~(\ref{eq:fv}). Increasing this to, e.g., $10^5$ would require a 10-fold
reduction in $\delta t$ and a much longer computation; however, since there are
$\approx 10^4$ steps/revolution, some runs already exceed $10^7$ steps, so this
is not presently feasible in general.

The relatively small cylinder diameters and fill levels used here (details
below) limit the compression forces that particles can experience. At the base
of a static column of 10 small particles the overlap (which can be regarded as
compression) will be 5\% (for $k_n$ and $g$ as given), an amount that is
unlikely to alter the qualitative behavior; indeed, tangential forces will be
enhanced for softer particles that experience a greater number of longer-lasting
contacts. (Particles with the $r^{-12}$ repulsion that is part of the force used
in \cite{rap02} have an overlap three times larger; the ultimately greater
resistance of this force to compression asserts itself only at five times the
depth.) Systems with substantially thicker layers will require a stiffer volume
interaction (and consequently a smaller $\delta t$).

The values of the remaining parameters from \cite{rap02} are as follows. In
Eq.~(\ref{eq:fd}), $\gamma_n = 5$. The coefficients in Eq.~(\ref{eq:fs}) are
$\gamma_s^{bb} = 10$, and in general $\gamma_s^{ss} = \gamma_s^{bs} = 2$, except
for a few cases where all $\gamma_s = 10$; the particle-wall values are
identical. The relative values of the static friction coefficients, e.g.,
$\mu^{bb} / \mu^{ss}$, are set equal to the ratio of the corresponding
$\gamma_s$ values, with the larger of the pair equal to 0.5. Finally, the value
of the new parameter introduced with Eq.~(\ref{eq:udef}) is $k_g = 500$,
although, as with most parameter settings, there is considerable latitude, and
all that is required at this exploratory stage is for the results to appear
qualitatively reasonable.

The nominal particle diameter is the interaction cutoff in Eq.~(\ref{eq:fv}).
For small particles, to maintain consistency with \cite{rap02}, $d_s = 2^{1/6}
\approx 1.122$ (MD units), while for big particles $d_b = b d_s$. The actual
particle sizes are uniformly distributed over a narrow range $[ d_s - 0.2, d_s
]$ for small particles, and likewise for big; the mean diameter of the small
particles is then close to unity. Particles all have the same density, so the
big to small mass ratio (before allowing for the random size distribution) is
$b^3$. The relative population of big and small particles is chosen to give
equal volume fractions; for most runs $b = 1.8$, with a big particle fraction of
0.15.

The cylinder diameter $D$ and length $L$ have value ranges 30--40 and 120--360
(MD units) respectively. (A consequence of the use of soft potentials is that
the effective $D$ and $L$ values are reduced by approximately one unit.) The
aspect ratio $L / D$, a quantity having some influence on the number of axial
segregation bands \cite{cha06}, lies between 4 and 12. The cylinders used here
are relatively narrow, with $D$ less than 30 times the mean particle diameter, a
limitation that has been found experimentally to influence behavior
\cite{ale04}; even so, the number of particles in a simulation can exceed
70\,000. Another connection between experiment and simulation is the
dimensionless Froude number ${\rm Fr} = \Omega^2 D / 2 g$, the ratio of
centrifugal to gravitational acceleration. Experimentally ${\rm Fr} \approx
10^{-3}$--$10^{-1}$, and to ensure the simulations remain in the correct regime,
the value should not be allowed to become too large; this sets an upper limit to
$\Omega$, and for the simulations ${\rm Fr} < 0.2$.

\subsection{Measuring segregation}

Quantitative measures of the overall intensity of axial and radial segregation
are used to augment direct observation. The time-dependent axial segregation,
$S_a(t)$, is evaluated from binned counts along the axis weighted by particle
mass,
\begin{equation}
S_a = \frac{ \langle | b^3 n_b(q) - n_s(q) | \rangle_q } { b^3 n_b + n_s } ,
\label{eq:sa}
\end{equation}
where $n_b(q)$ and $n_s(q)$ are the number of big and small particles in a slice
(width $\approx 2.5$) centered at $q$ along the axis, $\langle \ldots \rangle_q$
denotes an average over all slices, and $n_b$ and $n_s$ are the totals. Binned
counts are less suitable for evaluating the radial segregation, $S_r(t)$,
because, if done in the same way as the space-time plots discussed below, not
all particles would be able to contribute. An alternative measure, based on the
mean-square radial distance of each type of particle from its center of mass, is
therefore used,
\begin{equation}
S_r = \frac{ ( \langle r_b^2 \rangle - \langle r_b^{\vphantom{2}} \rangle^2 ) -
( \langle r_s^2\rangle - \langle r_s^{\vphantom{2}} \rangle^2 ) }
{ \langle r^2 \rangle - \langle r \rangle^2 } , \label{eq:sr}
\end{equation}
where $\langle r_b \rangle$ and $\langle r_s \rangle$ denote the mean distances
of big and small particles from the axis, and $\langle r \rangle$ the overall
mean. Time is expressed in terms of the number of cylinder revolutions, $n_R$.

\section{Results}

Due to the very nature of the segregation effect, not all of its features are
readily quantifiable. The functions $S_a(t)$ and $S_r(t)$,
Eqs.~(\ref{eq:sa}-\ref{eq:sr}), provide a global summary of pattern development,
but lack detailed information concerning the number of axial bands, the
sharpness of the boundaries between segregated regions and the regularity of the
band pattern. The use of space-time plots, coded using either color or grayscale
level, provides a much more detailed picture of how both axial and radial
segregation evolve with time. Such plots show the relative populations, weighted
by particle volumes, in appropriately oriented slices. For the axial plots,
slices are normal to the cylinder axis; while this does not provide the same
information as the surface populations seen in experiment, it will be similar
provided that radial inhomogeneity is weak. For the radial plots, slices are
parallel to the upper free surface of the material; the nominal surface slope is
determined by a linear fit to the inner 2/3 of the surface away from the curved
boundary, and only the particle populations in a slab of similar width normal to
the surface are counted, to avoid bias due to the shape of the region.

Even more detailed information can be extracted from snapshots of the entire
system, in which particular subsets of particles can be selected for viewing;
images of this kind can, for example, reveal interior organization analogous to
that observed experimentally with MRI. An even richer visual approach employs
animated recordings showing the full temporal development of the system. An
animation of this type consists of an extended series of snapshots, recorded at
regular intervals throughout the run; all the information needed for the
analysis that follows is in fact obtained from configurations reconstructed from
such recordings.

The runs described here are cataloged in Table I; they are labeled
alphabetically for reference and listed in the order they first appear in the
discussion. Run lengths were not specified in advance; a run was generally
allowed to continue for as long as something interesting was happening, or until
it seemed that the system had stabilized (the possibility of premature
termination can never be ruled out), or until changes appeared to be occurring
too slowly to warrant continuation.

\begingroup
\squeezetable
\begin{table}
\caption{\label{tab:runs} Summary of runs discussed in the text; the top row
includes the default settings used unless indicated otherwise.}
\begin{ruledtabular}
\begin{tabular}{lcccccccccclrcc}
Id\footnote{Runs are denoted in the text as \#{\em A}, etc.} &
\multicolumn{6}{c}{Size, etc.\footnote{Cylinder length $L$ and diameter $D$,
filling density $\rho$, number of particles $N$ (rounded to nearest 100), big
particle size $b$, angular velocity $\Omega$.}} &
M\footnote{M/S: Mixed or segregated initial state.} &
\multicolumn{3}{c}{Forces\footnote{Values of force coefficients: $k_n$,
$k_g$ and $\gamma_s^{ss}$; all others are constant with values listed in
the text.}} &
R\footnote{R denotes repeat of preceding run with different initial state.} &
\multicolumn{3}{c}{Outcome} \\
& $L$ & $D$ & $\rho$ & $N$ & $b$ & $\Omega$ & & $k_n$ & $k_g$ &
$\gamma_s^{ss}$ & & $n_R$\footnote{Run length, in cylinder revolutions.} &
Rd\footnote{Radial segregation: permanent, transient, inverted, or none.} &
Ax\footnote{Axial segregation: initial and final band count, or none.} \\
\hline
{\em A} &240&40& .30& 71\,600 &1.8&.2&M&1000&500& 2&  & 1620 & P & 12        \\
{\em B} &360&30&    & 57\,600 &   &.1& &    &   &  &  & 2630 & T & 17$\to$11 \\
{\em C} &160&40&    & 47\,300 &   &.1& &    &   &  &  & 1800 & P & 7         \\
{\em D} &120&30& .50& 35\,900 &   &  & &    &   &  &  &  470 & P & -         \\
{\em E} &120&30&    & 18\,800 &   &  & &    &   &  &  & 2940 & T & 6         \\
{\em F} &120&30&    & 18\,800 &   &  & &    &   &  &R & 1310 & T & 5         \\
{\em G} &120&30&    & 18\,800 &   &  & &    &   &  &R & 1310 & T & 7$\to$4   \\
{\em H} &120&30& .40& 24\,200 &   &  & &    &   &  &  &  730 & P & -         \\
{\em I} &120&30& .45& 29\,300 &   &  & &    &   &  &  &  590 & P & -         \\
{\em J} &120&30&    & 18\,800 &   &.1& &    &   &  &  & 4290 & T & 6$\to$3   \\
{\em K} &360&30&    & 57\,600 &   &  & &    &   &  &  & 2960 & T & 17$\to$12 \\
{\em L} &160&40&    & 47\,300 &   &  & &    &   &  &  & 3320 & P & 7         \\
{\em M} &120&30&    & 18\,800 &   &  & &    &  0&  &  & 2040 & I & 3         \\
{\em N} &120&30&    & 18\,800 &   &  & &    &  0&  &R & 1030 & I & 5         \\
{\em O} &120&30& .45& 29\,300 &   &  & &    &  0&  &  &  830 & P & -         \\
{\em P} &120&30&    & 19\,200 &1.5&  & &    &   &  &  & 6350 & - & 7$\to$5   \\
{\em Q} &120&30&    & 19\,200 &1.3&  & &    &   &  &  & 2760 & - & 6         \\
{\em R} &120&30&    & 18\,800 &   &  & &2000&   &  &  & 2610 & T & 6$\to$3   \\
{\em S} &120&30&    & 18\,800 &   &  & &    &   &10&  & 6350 & P & 
  2\footnote{Partial segregation only.}\\
{\em T} &180&30&    & 16\,800 &   &  &S&    &   &  &  & 2840 & - & 2         \\
{\em U} &180&30&    & 16\,800 &   &  &S&    &   &10&  & 2440 & P & -         \\
{\em V} &160&40& .18& 27\,900 &   &  &S&    &   &  &  & 6110 & - & 2$\to$4   \\
\end{tabular}
\end{ruledtabular}
\end{table}
\endgroup

Most runs are begun from a uniformly mixed initial state, and in a few cases,
from a state that is presegregated into two axial bands. The particles are
placed on a lattice and assigned small random velocities (details that vanish
after just a few collisions); particle species is either randomly assigned
according to relative concentration, or, in the presegregated case, determined
by axial position. A change of random number seed allows repeated runs of
systems that are otherwise identical.

In all cases the cylinder is capped at the ends. Use of an axially periodic
cylinder leads to similar pattern development, establishing that caps are not
required for segregation. However, as in \cite{rap02}, the entire band structure
is then subject to axial drift. Since there is no evidence that the caps affect
the results (apart from relaxing the requirement for an even number of bands in
the periodic case) they can be used without concern.

In view of the extensive simulations involved, it is interesting to consider the
computational performance. For a system of 47\,000 particles, and a partly
parallelized computation running on a dual 3.6\,GHz Intel Xeon workstation, the
simulation proceeds at a rate of approximately 23\,000 time steps/hr, equivalent
to 3.7 revolutions/hr at $\Omega = 0.2$; some of the runs therefore extend over
several weeks. Contrast this with a typical experiment that can be run at, e.g.,
1000 revolutions/hr, irrespective of the number of particles.

\subsection{Axial and radial segregation}

The first two runs discussed are among the more extensive carried out. While
similar in respect to the initial radial segregation that peaks after some
10--20 revolutions and the appearance of multiple axial bands, in one instance
radial segregation persists while in the other it practically vanishes. The
details of these two runs, \#{\em A} and \#{\em B}, will be described using
axial and radial space-time plots, together with pictures of the final states
that reveal the nature of the internal structure, and with graphs of segregation
as measured by $S_a$ and $S_r$. Both runs suggest that a steady final state has
been achieved, although, given the experimentally observed long-term slowdown in
the bond merging rate, there is no way of completely excluding future changes of
this type, no matter how long the run.

\begin{figure}
\includegraphics[scale=1.80]{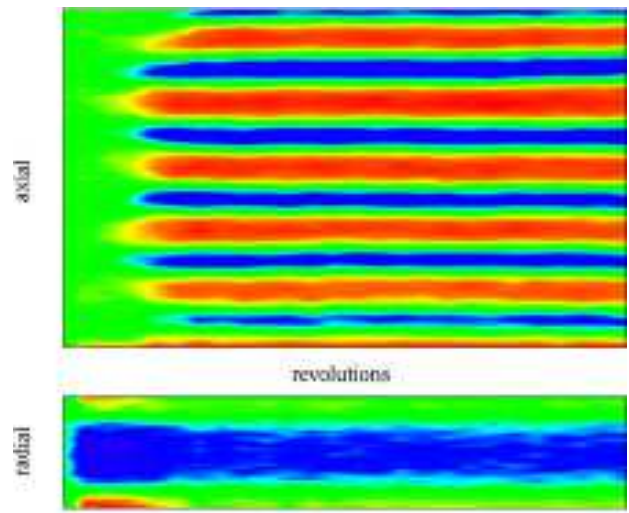}
\caption{\label{fig:f01} (Color online) Axial and radial space-time plots for
run \#{\em A} ($n_R = 1620$, $L = 240$, $L / D = 6$); red and blue (or medium
and dark gray) denote higher volume fractions of big or small particles. Time,
expressed as cylinder revolutions, is along the horizontal axis, and the
vertical axis measures axial or radial (the latter normal to the free surface)
position.}
\end{figure}

Fig.~\ref{fig:f01} shows the axial and radial space-time plots for run \#{\em
A}. Radial segregation appears very early in the run and a core of small
particles persists throughout, although the outer layer rich in big particles
disappears from the radial plot once axial banding begins; after the 12 axial
bands have formed the pattern appears stable, with no hint of any future change.

\begin{figure}
\includegraphics[scale=1.75]{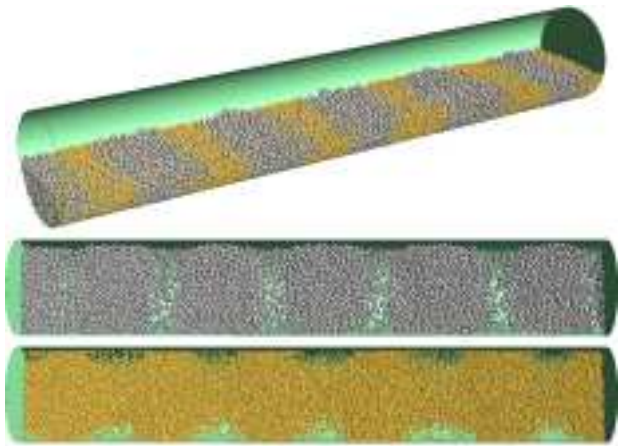}
\caption{\label{fig:f02} (Color online) Final state of run \#{\em A}; the full
system, and views showing just the big and small particles (colored silver and
gold).}
\end{figure}

Images of the final state of run \#{\em A} appear in Fig.~\ref{fig:f02}. The
first is an oblique view of the full system showing the 12 alternating bands.
The others are views looking down in a direction normal to the surface that show
big and small particles separately. These interior views reveal that while the
bands of big particles are not entirely separate, there are visible divisions
between them, and that the bands of small particles are joined by a central core
extending the length of the system.

\begin{figure}
\includegraphics[scale=1.80]{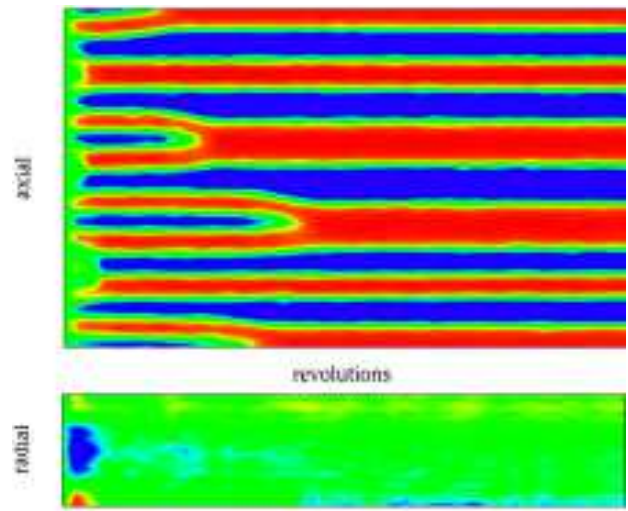}
\caption{\label{fig:f03} (Color online) Axial and radial space-time plots for
\#{\em B} ($n_R = 2630$, $L = 360$, $L / D = 12$).}
\end{figure}

Fig.~\ref{fig:f03} shows the space-time plots for run \#{\em B}. Here, unlike
\#{\em A}, radial segregation is a transient effect lasting less than 200
revolutions; furthermore, the early axial pattern is not maintained, and the
initial 17 bands are eventually reduced to 11 due to the vanishing of the small
particle bands at the ends and the merging of two pairs of big particle bands
(eliminating two more small particle bands). During the latter portion of the
run, extending over more than half its total length, there is no suggestion of
further pattern change. Band coarsening is a well-known experimental result,
with the band count falling roughly logarithmically with time \cite{fie03},
although the dependence can be more complicated than this \cite{fin06}.

\begin{figure}
\includegraphics[scale=1.75]{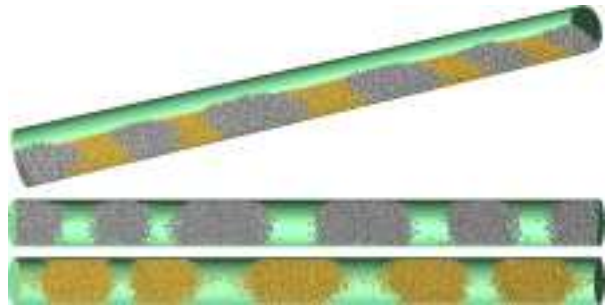}
\caption{\label{fig:f04} (Color online) Final state of \#{\em B}.}
\end{figure}

The images in Fig.~\ref{fig:f04} show the final state of \#{\em B}. Band
separation is practically complete in this case, with essentially no misplaced
big particles, and only the faintest remnant of a small particle core.

\begin{figure}
\includegraphics[scale=0.85]{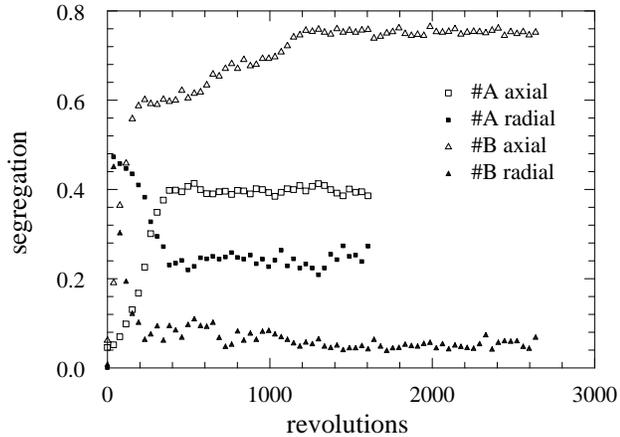}
\caption{\label{fig:f05} Time-dependent axial and radial segregation, $S_a$ and
$S_r$, for runs \#{\em A} and \#{\em B}; time is expressed in cylinder
revolutions.}
\end{figure}

The development of axial and radial segregation, $S_a$ and $S_r$, for these two
runs is shown in Fig.~\ref{fig:f05}. The graphs reflect what has already been
noted in the space-time plots and images, namely, that a preference for small
particles in the core persists over the duration of \#{\em A}, but not \#{\em
B}; the presence of this core affects the magnitude of $S_a$, which is smaller
in \#{\em A} than in \#{\em B}.

\begin{figure}
\includegraphics[scale=1.50]{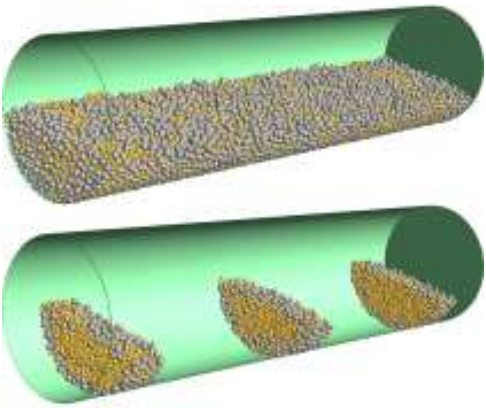}
\caption{\label{fig:f06} (Color online) Views of \#{\em C} after 60 revolutions;
the full system and three slices are shown.}
\end{figure}

The images in the next two figures provide examples of radial segregation in
cases where the effect is well developed.
Fig.~\ref{fig:f06} shows \#{\em C} after 60 revolutions;
this run eventually develops axial segregation. The complete system is shown,
together with three narrow slices (of thickness $0.03 L$) at the midpoint and at
a distance $0.1 L$ from either end.

\begin{figure}
\includegraphics[scale=1.05]{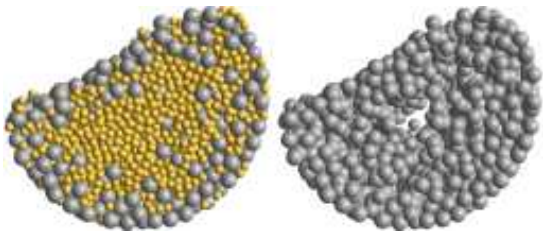}
\caption{\label{fig:f07} (Color online) Views of \#{\em D} after 260 revolutions
(see text for explanation).}
\end{figure}

Fig.~\ref{fig:f07} shows \#{\em D} after 260 revolutions (graphs of $S_a$ and
$S_r$ for this run appear later). The view is along the
axis, after slices of thickness $0.1 L$ are removed from each end; on the left
all particles are shown, while on the right just the big particles appear. There
is a visible opening extending along the entire length ($0.8 L$) near the center
of the small particle core; while the simulations do not produce a totally pure
core, radial segregation is strong despite misplaced big particles; this rather
short run showed no hint of imminent axial segregation.

Two general observations concerning segregation emerge from these (and
subsequent) simulation runs. When radial segregation occurs it takes the form of
a core of small particles surrounded by big particles (there is an exception
that is discussed below); this is also the case experimentally, although in the
simulations the core boundaries are not as sharp and the effect appears to
develop more slowly. There is no preferred particle type near the end caps nor a
tendency for bands to nucleate there; bands can form at different axial
positions at different times; sometimes bond formation is almost simultaneous
and sometimes not.

\subsection{Reproducibility}

The issue of reproducibility is of particular importance since conclusions must
take into account the variability of the behavior, such as different
intermediate or terminal band counts. Multiple runs would be required, as is
normal experimentally. In most cases this condition has not been fulfilled, but
an example of three runs involving systems that are identical, apart from the
initial random state, is considered here (and one further example appears
later).

\begin{figure}
\includegraphics[scale=1.80]{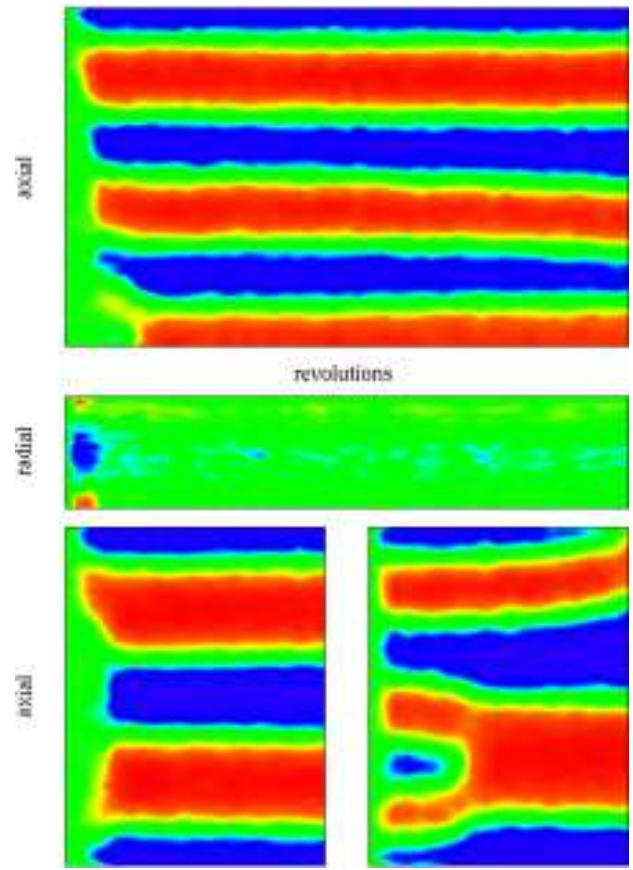}
\caption{\label{fig:f08} (Color online) Axial and radial space-time plots for
run \#{\em E} ($n_R = 2940$) and, on the same scale, axial plots for\#{\em F}
and \#{\em G} ($n_R = 1310$) with different initial states ($L = 120$, $L / D =
4$).}
\end{figure}

Fig.~\ref{fig:f08} shows the axial and radial space-time plots for run \#{\em
E}, together with the axial plots for the shorter runs \#{\em F} and \#{\em G}.
The behavior differs (although $S_a$ and $S_r$ are only weakly affected);
each exhibits early radial segregation, but
they differ in regard to axial band development. This is just what happens
experimentally under appropriate conditions \cite{fie03}; some aspects of the
segregation ($S_a$ and $S_r$) are reasonably robust, while others (the band
details) may vary significantly between runs. A preferred wavelength governing
the axial band pattern could exist, but since development history can influence
outcome, it is probably only meaningful to refer to a mean number of bands (and
a mean wavelength).

\subsection{Parameter dependence}

When comparing simulation to experiment, the parameters defining the system fall
into two categories. There are those that are readily changed in an experiment,
such as the cylinder rotation rate, the fill level, and even the cylinder size;
experimental results may cover a range of these parameters in order to
demonstrate trends or construct phase diagrams. The others are related to the
properties of the particles themselves; while there are numerous types of
granular material available for experimental study, efforts have tended to focus
on a rather limited subset, so that a systematic coverage of behavior in terms
of material properties is presently unavailable.

\begin{figure}
\includegraphics[scale=0.85]{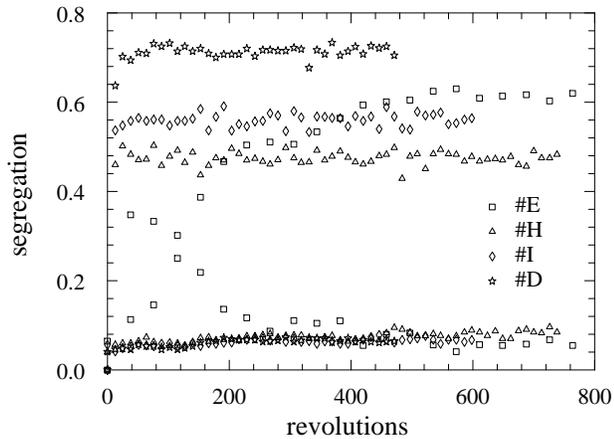}
\caption{\label{fig:f09} Axial and radial segregation (the graphs for the latter
peak early in each case) for runs \#{\em E}, \#{\em H}, \#{\em I} and \#{\em D}
with increasing fill level.}
\end{figure}

Changing the fill level can alter the nature of the segregation, as seen
experimentally with slurries \cite{arn05}. Fig.~\ref{fig:f09} shows $S_a$ and
$S_r$ for four runs differing only in their fill level, \#{\em E}, \#{\em H},
\#{\em I} and \#{\em D} (because of the slight reduction in effective cylinder
size mentioned earlier, $\rho = 0.45$ corresponds to a cylinder that is
approximately half full, while for $\rho = 0.3$ the layer depth is close to
$\frac{3}{8} D$). In \#{\em E}, transient radial segregation is followed by
axial, while in the other three cases only radial segregation occurs. These and
other runs suggest that a sufficiently high fill level favors permanent radial
segregation and at the same time suppresses axial segregation. Validation of the
model, and determining how closely it reproduces experiment, requires that
behavioral trends under a change of parameters be reproduced correctly; while
experiment suggests band suppression at low fill levels \cite{arn05}, this
applies to slurries, and the degree of similarity between dry and wet mixtures
has yet to be studied systematically.

The cylinder angular velocity $\Omega$ is the parameter most readily changed in
an experiment. The $\Omega$ dependence (using the values $\Omega = 0.2$ and 0.1)
for three pairs of runs that are otherwise the same has been examined;
the run pairs are \#{\em E} and \#{\em J} that eventually
produce 6 and 3 bands, \#{\em K} and \#{\em B} that produce 12 and 11 bands and,
lastly, \#{\em L} and \#{\em C} that both produce 7 bands (a tentative result in
the case of \#{\em C}). A feature common to each of these pairs is that $S_a$ is
slightly larger at reduced $\Omega$.

The influence of other parameters will be mentioned only briefly. Runs \#{\em M}
and \#{\em N} show the effect of removing the tangential restoring force,
$k_g = 0$ in Eq.~(\ref{eq:udef}), in a sense making the particles less
``grain-like''. The runs are otherwise similar to \#{\em E} and, apart from the
function used for $\avec{f}_v$ in Eq.~(\ref{eq:fv}), now resemble \cite{rap02};
so do the results, in which axial segregation is now accompanied by inverse
radial segregation. This outcome confirms the importance of $\avec{f}_g$ for
obtaining the correct radial behavior. Early in these runs weak normal radial
segregation occurs, but it becomes inverted when the axial bands appear.
Different behavior is encountered at a higher fill level; run \#{\em O}, which
is related to run \#{\em I} in a similar way, shows normal radial but no axial
segregation. Outcomes such as these hint at the complex interplay of parameters
and call for an extensive coverage, with multiple runs, prior to drawing
conclusions.

Another key parameter is $b$, the relative size of the big particles. In run
\#{\em P}, $b$ is reduced to 1.5, and 5 axial bands appear, but no measurable
radial segregation; $b$ is further reduced to 1.3 (the preferred value in
\cite{rap02}) in run \#{\em Q}, resulting in 6 bands. In the opposite direction,
it is questionable how large $b$ can become without having to increase cylinder
diameter to avoid serious size effects (and, to maintain aspect ratio, also its
length), but for $b = 2.1$ (not shown), depending on fill level, both radial and
axial segregation can be obtained; experiment typically involves $b \approx 3$
or greater.

Making the particles slightly less compressible by doubling $k_n$, in run \#{\em
R}, leads to initial radial segregation, and eventually three axial bands.
A more substantial increase in $k_n$ would, as noted earlier, require a smaller
$\delta t$. Further runs (not shown) similar to \#{\em A}, but with $k_n$ and
all the other interaction coefficients increased by factors of 10 or 100, and
with suitably reduced $\delta t$, produce the same segregation effects,
confirming that this aspect of the model does not significantly influence the
behavior. A similar conclusion applies to granular chute flow \cite{sil01}.
For a final example of parameter dependence (requiring further
study), run \#{\em S} is similar to \#{\em E} except that $\gamma_s^{ss} =
\gamma_s^{bb} = 10$; radial segregation occurs, suggesting that this effect can
be driven by particle size difference alone, and eventually two weakly developed
axial bands appear (the outcome of this long run is less clear than usual
since a final state had yet to develop).

\subsection{Particle motion}

The configurational snapshots recorded during the runs enable post-run analysis
of the motion of individual or selected groups of particles over extended
intervals. An eventual goal for this kind of analysis is developing a capability
for relating local organization and dynamics at the level of individual
particles to the behavior at a scale where the collective nature of segregation
is exhibited. Two examples will be considered here.

\begin{figure}
\includegraphics[scale=1.75]{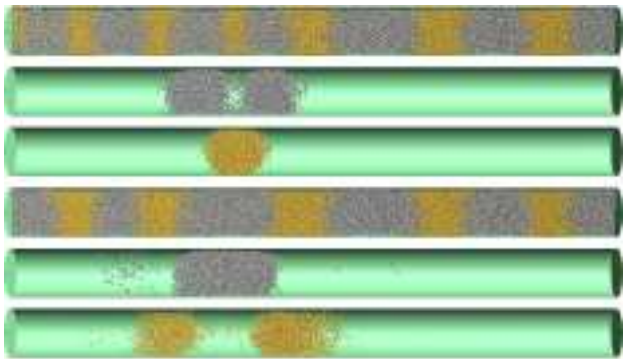}
\caption{\label{fig:f10} (Color online) Band merging in run \#{\em B} where
small particles from the middle band disperse allowing the big particle bands to
merge; the views show the full system and the particles from the bands involved,
both before and after the event.}
\end{figure}

The first is the merging of two bands of big particles, accompanied by the
dispersal of the small particles from the disappearing middle band. A merge
event of this type occurs in run \#{\em B} (see Fig.~\ref{fig:f03}) between
revolutions 700 and 1200. The upper portion of Fig.~\ref{fig:f10} shows three
views prior to the event, namely the full system, and the selected bands of big
and small particles (the latter also revealing small particles that intrude into
the big particle band, a detail hidden from the outside). The lower portion
shows three views at the completion of the event, namely the full system, and
all the previously selected particles in their new positions, principally within
the bands directly involved. Even though there is essentially no residual
small particle core at this stage of the run, small particles are able to
migrate across the big particle bands more readily than the converse; selective
migration of this kind will be even more apparent in the next example.
An interesting feature revealed by these images is that while the small
particles are dispersed in both directions, more seem to have traveled to the
right where the original band of big particles was somewhat narrower; while such
behavior might not be unexpected it is gratifying to see it actually occurring.

\begin{figure}
\includegraphics[scale=1.75]{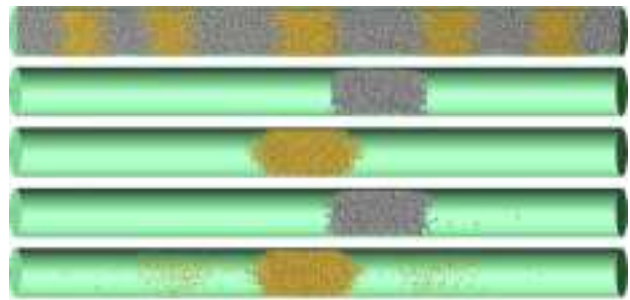}
\caption{\label{fig:f11} (Color online) Particle migration and confinement over
an interval of almost 1400 revolutions during run \#{\em B}; the views show the
full system at the start of the interval together with the selected bands of big
and small particles, and the same sets of big and small particles at the end.}
\end{figure}

The second example also involves run \#{\em B}, but now for monitoring the
steady state over a relatively long interval between revolutions 1230 and 2600.
The first view in Fig.~\ref{fig:f11} is of the entire system at the beginning;
the next two views show the adjacent, slightly overlapped bands of big and small
particles selected for tracking. The last two views show all the selected
particles in their final positions (since the band structure is stationary,
views of the entire system during the interval are practically
indistinguishable); although quite a few small particles (but constituting only
a tiny fraction of the band population) have migrated out of the band in both
directions, almost no big particles have done likewise, demonstrating that they
are effectively confined for lack of an escape path. Departing
particles are replaced by others, but these are not shown. This form of analysis
can also be carried out experimentally by adding tracer particles to existing
bands \cite{fin06}, revealing that confinement is indeed a strong effect.

\subsection{Presegregated systems}

\begin{figure}
\includegraphics[scale=1.50]{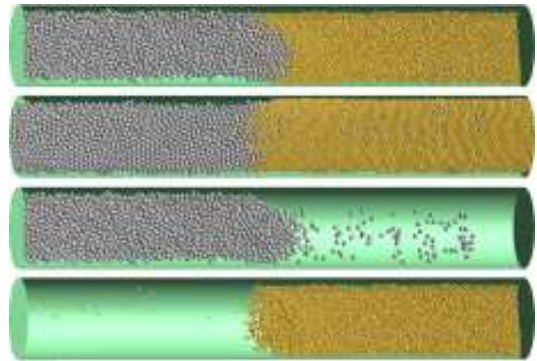}
\caption{\label{fig:f12} (Color online) Final state of \#{\em T}; top, bottom,
big and small particle views.}
\end{figure}

An alternative perspective is provided by systems that begin in an axially
segregated state, a problem also studied experimentally with MRI
\cite{rna99,fin06}. Various scenarios exist, of which preserving the original
state and axial mixing represent two extremes, while rearrangement into some
other band pattern is also a possibility. Examples of all three types
will be considered. Very little change occurs over the duration of run \#{\em T}.
Fig.~\ref{fig:f12} shows its final state, and from the
top and bottom views it is apparent that the interface has adopted a curved
form. The views showing big and small particles separately reveal that
comparatively small numbers of each type have penetrated into the opposite
region; interestingly, in this example big particles also appear to be
participating in the migration.

\begin{figure}
\includegraphics[scale=1.80]{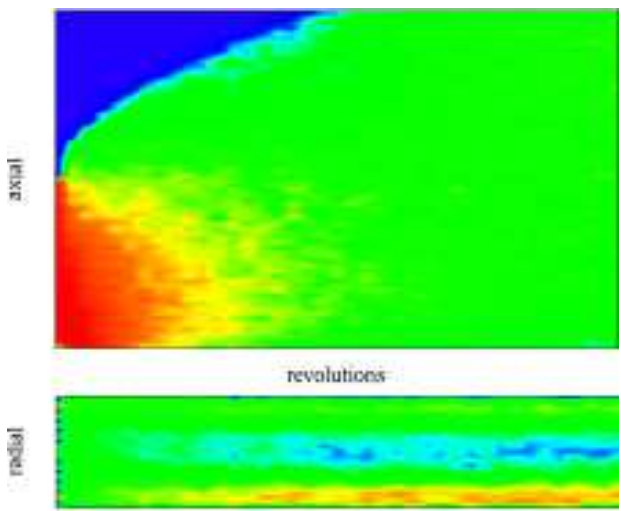}
\caption{\label{fig:f13} (Color online) Axial and radial space-time plots for
\#{\em U} ($n_R = 2440$, $L = 180$, $L / D = 6$).}
\end{figure}

Mixing is most easily achieved by setting all the friction parameters equal (as
in \#{\em S}, where only a weak axial effect was seen).
Fig.~\ref{fig:f13} shows the outcome of this in run \#{\em U};
the initial axial segregation vanishes promptly and a certain amount of radial
segregation appears.

\begin{figure}
\includegraphics[scale=1.80]{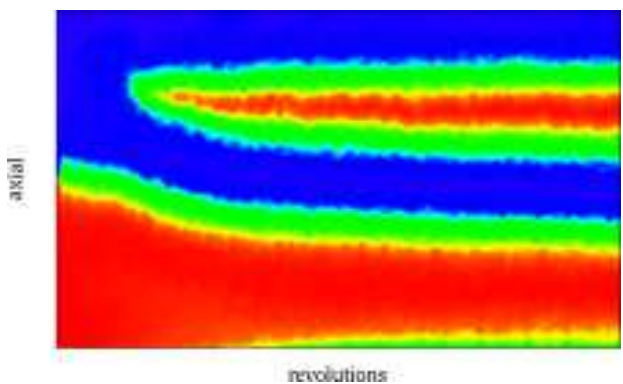}
\caption{\label{fig:f14} (Color online) Axial space-time plot for \#{\em V}
($n_R = 6110$, $L = 160$, $L / D = 4$).}
\end{figure}

The final scenario, the appearance of an altered axial band pattern, occurs in
run \#{\em V} (similar to \#{\em L} but with fewer particles), shown in
Fig.~\ref{fig:f14}. The most prominent feature here is the splitting of the
small particle band by a new band of big particles. The process involves
particle migration in both directions; from the recorded configurations (not
shown) it appears that small particles follow an interior path while big
particles tend to prefer a path close to the curved cylinder wall; the dynamics
of this process is yet another aspect of the overall problem needing further
study.

\subsection{Surface profiles}

The motion of particles along the upper free surface plays a role in the
segregation process, although the importance of this contribution and the
mechanism involved is open to question. The detailed form of the surface is
therefore of interest, although it is not readily characterized and, for a
specific particle mixture, it will depend on fill level, rotation rate and
cylinder diameter. The profile of the free surface typically varies slightly
with position along the axis once axial segregation has occurred. For these
reasons, just one instance is included here. Experimental examples, ranging from
strongly S-shaped to almost linear, appear in \cite{lev99}.

\begin{figure}
\includegraphics[scale=0.9]{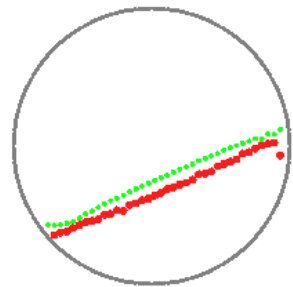}
\caption{\label{fig:f15} Averaged surface profiles for big and small particles
(big and small points) for run \#{\em L} ($D = 40$).}
\end{figure}

Fig.~\ref{fig:f15} shows the axially averaged surface profile near the end of
run \#{\em L}, where there are seven bands; big and small particles are treated
individually, and the results represent an average over ten configurations
spaced six revolutions apart. The profiles are reasonably close to linear
(curved profiles similar to \cite{rap02} can be obtained at lower fill levels);
the big particles appear slightly below the small because it is the particle
centers that are monitored. There are deviations adjacent to the curved cylinder
wall, but here the data binning produces some distortion. Whether the slight
deviations from linearity are important for the segregation process remains to
be seen. Except at the start, there is no obvious change in this profile over
the entire run.

\section{Conclusion}

Simulations based on a simple particle-based model of granular matter have been
used to demonstrate the occurrence of both radial and axial segregation in a
rotating cylinder containing a two-component granular mixture. The earlier
inability to reproduce radial segregation correctly has been resolved; for the
type of model considered, this is due to the omission of a component of the
frictional force that helps resist sliding during collisions. The different
forms of segregation observed are indicative of a large, complex parameter
space, as is the case experimentally.

In order to learn more about the phase diagram and determine which parameters
dominate in different regimes, a systematic study, including multiple
realizations due to reproducibility issues, is required. The results suggest
many questions of the ``what if'' type; of necessity, the answers will also have
to await further work. The model itself is open to critical examination, in
particular, how the components of the interparticle force influence the behavior
(one example has already been given), as well as the relevant ranges of the
force parameters. More generally, in view of the way static friction is
represented, how reliable is the model altogether (a question equally applicable
to similar models appearing elsewhere)?

In theoretical fluid dynamics, stability analysis \cite{cha61} is used to
determine when symmetry breaking is advantageous. No analogous theory exists for
the granular segregation phenomena described here,
and the question of what distinguishes the different
segregation modes from one another, and from the uniformly mixed state, remains
open; is it, for example, the ability to optimize energy dissipation, thereby
ensuring the most efficient flow? The segregation band coarsening slows with
increasing band width, and the states eventually reached have varying numbers of
bands and degrees of pattern regularity; here the question is whether such
patterns are stable, or are they slow transients en route to even further
coarsening? Another issue is the relation between the radial and axial
segregation processes; are they independent phenomena that just happen to
coexist (perhaps in competition) in most, but not all situations, or does the
former actually drive the latter? Finally, no attempt has been made to probe the
underlying mechanisms beyond the exercises in particle tracking; if simulation
can be shown to reproduce experiment reliably, then, in view of the level of
detail provided, it may prove an important tool for exploring the complexities
of segregation.

\bibliography{grancyl}

\begin{thebibliography}{39}
\expandafter\ifx\csname natexlab\endcsname\relax\def\natexlab#1{#1}\fi
\expandafter\ifx\csname bibnamefont\endcsname\relax
  \def\bibnamefont#1{#1}\fi
\expandafter\ifx\csname bibfnamefont\endcsname\relax
  \def\bibfnamefont#1{#1}\fi
\expandafter\ifx\csname citenamefont\endcsname\relax
  \def\citenamefont#1{#1}\fi
\expandafter\ifx\csname url\endcsname\relax
  \def\url#1{\texttt{#1}}\fi
\expandafter\ifx\csname urlprefix\endcsname\relax\def\urlprefix{URL }\fi
\providecommand{\bibinfo}[2]{#2}
\providecommand{\eprint}[2][]{\url{#2}}

\bibitem[{\citenamefont{Barker}(1994)}]{bar94}
\bibinfo{author}{\bibfnamefont{G.~C.} \bibnamefont{Barker}}, in
  \emph{\bibinfo{booktitle}{Granular Matter: An Interdisciplinary Approach}},
  edited by \bibinfo{editor}{\bibfnamefont{A.}~\bibnamefont{Mehta}}
  (\bibinfo{publisher}{Springer}, \bibinfo{address}{Heidelberg},
  \bibinfo{year}{1994}), p.~\bibinfo{pages}{35}.

\bibitem[{\citenamefont{Herrmann}(1995)}]{her95}
\bibinfo{author}{\bibfnamefont{H.~J.} \bibnamefont{Herrmann}}, in
  \emph{\bibinfo{booktitle}{3rd Granada Lectures in Computational Physics}},
  edited by \bibinfo{editor}{\bibfnamefont{P.~L.} \bibnamefont{Garrido}}
  \bibnamefont{and} \bibinfo{editor}{\bibfnamefont{J.}~\bibnamefont{Marro}}
  (\bibinfo{publisher}{Springer}, \bibinfo{address}{Heidelberg},
  \bibinfo{year}{1995}), p.~\bibinfo{pages}{67}.

\bibitem[{\citenamefont{Jaeger et~al.}(1996)\citenamefont{Jaeger, Nagel, and
  Behringer}}]{jae96}
\bibinfo{author}{\bibfnamefont{H.~M.} \bibnamefont{Jaeger}},
  \bibinfo{author}{\bibfnamefont{S.~R.} \bibnamefont{Nagel}}, \bibnamefont{and}
  \bibinfo{author}{\bibfnamefont{R.~P.} \bibnamefont{Behringer}},
  \bibinfo{journal}{Rev. Mod. Phys.} \textbf{\bibinfo{volume}{68}},
  \bibinfo{pages}{1259} (\bibinfo{year}{1996}).

\bibitem[{\citenamefont{Kadanoff}(1999)}]{kad99}
\bibinfo{author}{\bibfnamefont{L.~P.} \bibnamefont{Kadanoff}},
  \bibinfo{journal}{Rev. Mod. Phys.} \textbf{\bibinfo{volume}{71}},
  \bibinfo{pages}{435} (\bibinfo{year}{1999}).

\bibitem[{\citenamefont{Walton}(1983)}]{wal83}
\bibinfo{author}{\bibfnamefont{O.~R.} \bibnamefont{Walton}}, in
  \emph{\bibinfo{booktitle}{Mechanics of Granular Materials}}, edited by
  \bibinfo{editor}{\bibfnamefont{J.~T.} \bibnamefont{Jenkins}}
  \bibnamefont{and} \bibinfo{editor}{\bibfnamefont{M.}~\bibnamefont{Satake}}
  (\bibinfo{publisher}{Elsevier}, \bibinfo{address}{Amsterdam},
  \bibinfo{year}{1983}), p. \bibinfo{pages}{327}.

\bibitem[{\citenamefont{Hirshfeld and Rapaport}(1997)}]{hir97}
\bibinfo{author}{\bibfnamefont{D.}~\bibnamefont{Hirshfeld}} \bibnamefont{and}
  \bibinfo{author}{\bibfnamefont{D.~C.} \bibnamefont{Rapaport}},
  \bibinfo{journal}{Phys. Rev. E} \textbf{\bibinfo{volume}{56}},
  \bibinfo{pages}{2012} (\bibinfo{year}{1997}).

\bibitem[{\citenamefont{Rosato et~al.}(1987)\citenamefont{Rosato, Strandburg,
  Prinz, and Swendsen}}]{ros87}
\bibinfo{author}{\bibfnamefont{A.}~\bibnamefont{Rosato}},
  \bibinfo{author}{\bibfnamefont{K.~J.} \bibnamefont{Strandburg}},
  \bibinfo{author}{\bibfnamefont{F.}~\bibnamefont{Prinz}}, \bibnamefont{and}
  \bibinfo{author}{\bibfnamefont{R.~H.} \bibnamefont{Swendsen}},
  \bibinfo{journal}{Phys. Rev. Lett.} \textbf{\bibinfo{volume}{58}},
  \bibinfo{pages}{1038} (\bibinfo{year}{1987}).

\bibitem[{\citenamefont{Gallas et~al.}(1996)\citenamefont{Gallas, Herrmann,
  P{\"o}schel, and Sokolowski}}]{gal96}
\bibinfo{author}{\bibfnamefont{J.~A.~C.} \bibnamefont{Gallas}},
  \bibinfo{author}{\bibfnamefont{H.~J.} \bibnamefont{Herrmann}},
  \bibinfo{author}{\bibfnamefont{T.}~\bibnamefont{P{\"o}schel}},
  \bibnamefont{and}
  \bibinfo{author}{\bibfnamefont{S.}~\bibnamefont{Sokolowski}},
  \bibinfo{journal}{J. Stat. Phys.} \textbf{\bibinfo{volume}{82}},
  \bibinfo{pages}{443} (\bibinfo{year}{1996}).

\bibitem[{\citenamefont{Rapaport}(2001)}]{rap01}
\bibinfo{author}{\bibfnamefont{D.~C.} \bibnamefont{Rapaport}},
  \bibinfo{journal}{Phys. Rev. E} \textbf{\bibinfo{volume}{64}},
  \bibinfo{pages}{061304} (\bibinfo{year}{2001}).

\bibitem[{\citenamefont{Cantelaube and Bideau}(1995)}]{can95}
\bibinfo{author}{\bibfnamefont{F.}~\bibnamefont{Cantelaube}} \bibnamefont{and}
  \bibinfo{author}{\bibfnamefont{D.}~\bibnamefont{Bideau}},
  \bibinfo{journal}{Europhys. Lett.} \textbf{\bibinfo{volume}{30}},
  \bibinfo{pages}{133} (\bibinfo{year}{1995}).

\bibitem[{\citenamefont{Zik et~al.}(1994)\citenamefont{Zik, Levine, Lipson,
  Shtrikman, and Stavans}}]{zik94}
\bibinfo{author}{\bibfnamefont{O.}~\bibnamefont{Zik}},
  \bibinfo{author}{\bibfnamefont{D.}~\bibnamefont{Levine}},
  \bibinfo{author}{\bibfnamefont{S.~G.} \bibnamefont{Lipson}},
  \bibinfo{author}{\bibfnamefont{S.}~\bibnamefont{Shtrikman}},
  \bibnamefont{and} \bibinfo{author}{\bibfnamefont{J.}~\bibnamefont{Stavans}},
  \bibinfo{journal}{Phys. Rev. Lett.} \textbf{\bibinfo{volume}{73}},
  \bibinfo{pages}{644} (\bibinfo{year}{1994}).

\bibitem[{\citenamefont{Rapaport}(2002)}]{rap02}
\bibinfo{author}{\bibfnamefont{D.~C.} \bibnamefont{Rapaport}},
  \bibinfo{journal}{Phys. Rev. E} \textbf{\bibinfo{volume}{65}},
  \bibinfo{pages}{061306} (\bibinfo{year}{2002}).

\bibitem[{\citenamefont{Hill et~al.}(1997{\natexlab{a}})\citenamefont{Hill,
  Caprihan, and Kakalios}}]{hil97s}
\bibinfo{author}{\bibfnamefont{K.~M.} \bibnamefont{Hill}},
  \bibinfo{author}{\bibfnamefont{A.}~\bibnamefont{Caprihan}}, \bibnamefont{and}
  \bibinfo{author}{\bibfnamefont{J.}~\bibnamefont{Kakalios}},
  \bibinfo{journal}{Phys. Rev. Lett.} \textbf{\bibinfo{volume}{78}},
  \bibinfo{pages}{50} (\bibinfo{year}{1997}{\natexlab{a}}).

\bibitem[{\citenamefont{Hill et~al.}(1997{\natexlab{b}})\citenamefont{Hill,
  Caprihan, and Kakalios}}]{hil97f}
\bibinfo{author}{\bibfnamefont{K.~M.} \bibnamefont{Hill}},
  \bibinfo{author}{\bibfnamefont{A.}~\bibnamefont{Caprihan}}, \bibnamefont{and}
  \bibinfo{author}{\bibfnamefont{J.}~\bibnamefont{Kakalios}},
  \bibinfo{journal}{Phys. Rev. E} \textbf{\bibinfo{volume}{56}},
  \bibinfo{pages}{4386} (\bibinfo{year}{1997}{\natexlab{b}}).

\bibitem[{\citenamefont{Nakagawa et~al.}(1997)\citenamefont{Nakagawa,
  Altobelli, Caprihan, and Fukushima}}]{nak97}
\bibinfo{author}{\bibfnamefont{M.}~\bibnamefont{Nakagawa}},
  \bibinfo{author}{\bibfnamefont{S.~A.} \bibnamefont{Altobelli}},
  \bibinfo{author}{\bibfnamefont{A.}~\bibnamefont{Caprihan}}, \bibnamefont{and}
  \bibinfo{author}{\bibfnamefont{E.}~\bibnamefont{Fukushima}},
  \bibinfo{journal}{Chem. Eng. Sci.} \textbf{\bibinfo{volume}{52}},
  \bibinfo{pages}{4423} (\bibinfo{year}{1997}).

\bibitem[{\citenamefont{Frette and Stavans}(1997)}]{fre97}
\bibinfo{author}{\bibfnamefont{V.}~\bibnamefont{Frette}} \bibnamefont{and}
  \bibinfo{author}{\bibfnamefont{J.}~\bibnamefont{Stavans}},
  \bibinfo{journal}{Phys. Rev. E} \textbf{\bibinfo{volume}{56}},
  \bibinfo{pages}{6981} (\bibinfo{year}{1997}).

\bibitem[{\citenamefont{Choo et~al.}(1997)\citenamefont{Choo, Molteno, and
  Morris}}]{cho97}
\bibinfo{author}{\bibfnamefont{K.}~\bibnamefont{Choo}},
  \bibinfo{author}{\bibfnamefont{T.~C.~A.} \bibnamefont{Molteno}},
  \bibnamefont{and} \bibinfo{author}{\bibfnamefont{S.~W.}
  \bibnamefont{Morris}}, \bibinfo{journal}{Phys. Rev. Lett.}
  \textbf{\bibinfo{volume}{79}}, \bibinfo{pages}{2975} (\bibinfo{year}{1997}).

\bibitem[{\citenamefont{Choo et~al.}(1998)\citenamefont{Choo, Baker, Molteno,
  and Morris}}]{cho98}
\bibinfo{author}{\bibfnamefont{K.}~\bibnamefont{Choo}},
  \bibinfo{author}{\bibfnamefont{M.~W.} \bibnamefont{Baker}},
  \bibinfo{author}{\bibfnamefont{T.~C.~A.} \bibnamefont{Molteno}},
  \bibnamefont{and} \bibinfo{author}{\bibfnamefont{S.~W.}
  \bibnamefont{Morris}}, \bibinfo{journal}{Phys. Rev. E}
  \textbf{\bibinfo{volume}{58}}, \bibinfo{pages}{6115} (\bibinfo{year}{1998}).

\bibitem[{\citenamefont{Stavans}(1998)}]{sta98}
\bibinfo{author}{\bibfnamefont{J.}~\bibnamefont{Stavans}}, \bibinfo{journal}{J.
  Stat. Phys.} \textbf{\bibinfo{volume}{93}}, \bibinfo{pages}{467}
  (\bibinfo{year}{1998}).

\bibitem[{\citenamefont{Fiedor and Ottino}(2003)}]{fie03}
\bibinfo{author}{\bibfnamefont{S.~J.} \bibnamefont{Fiedor}} \bibnamefont{and}
  \bibinfo{author}{\bibfnamefont{J.~M.} \bibnamefont{Ottino}},
  \bibinfo{journal}{Phys. Rev. Lett.} \textbf{\bibinfo{volume}{91}},
  \bibinfo{pages}{244301} (\bibinfo{year}{2003}).

\bibitem[{\citenamefont{Arndt et~al.}(2005)\citenamefont{Arndt,
  Siegmann-Hegerfeld, Fiedor, Ottino, and Lueptow}}]{arn05}
\bibinfo{author}{\bibfnamefont{T.}~\bibnamefont{Arndt}},
  \bibinfo{author}{\bibfnamefont{T.}~\bibnamefont{Siegmann-Hegerfeld}},
  \bibinfo{author}{\bibfnamefont{S.~J.} \bibnamefont{Fiedor}},
  \bibinfo{author}{\bibfnamefont{J.~M.} \bibnamefont{Ottino}},
  \bibnamefont{and} \bibinfo{author}{\bibfnamefont{R.~M.}
  \bibnamefont{Lueptow}}, \bibinfo{journal}{Phys. Rev. E}
  \textbf{\bibinfo{volume}{71}}, \bibinfo{pages}{011306}
  (\bibinfo{year}{2005}).

\bibitem[{\citenamefont{Finger et~al.}(2006)\citenamefont{Finger, Voigt,
  Stadler, Niessen, Naji, and Stannarius}}]{fin06}
\bibinfo{author}{\bibfnamefont{T.}~\bibnamefont{Finger}},
  \bibinfo{author}{\bibfnamefont{A.}~\bibnamefont{Voigt}},
  \bibinfo{author}{\bibfnamefont{J.}~\bibnamefont{Stadler}},
  \bibinfo{author}{\bibfnamefont{H.~G.} \bibnamefont{Niessen}},
  \bibinfo{author}{\bibfnamefont{L.}~\bibnamefont{Naji}}, \bibnamefont{and}
  \bibinfo{author}{\bibfnamefont{R.}~\bibnamefont{Stannarius}},
  \bibinfo{journal}{Phys. Rev. E} \textbf{\bibinfo{volume}{74}},
  \bibinfo{pages}{031312} (\bibinfo{year}{2006}).

\bibitem[{\citenamefont{Newey et~al.}(2004)\citenamefont{Newey, Ozik, van~der
  Meer, Ott, and Losert}}]{new04}
\bibinfo{author}{\bibfnamefont{M.}~\bibnamefont{Newey}},
  \bibinfo{author}{\bibfnamefont{J.}~\bibnamefont{Ozik}},
  \bibinfo{author}{\bibfnamefont{S.~M.} \bibnamefont{van~der Meer}},
  \bibinfo{author}{\bibfnamefont{E.}~\bibnamefont{Ott}}, \bibnamefont{and}
  \bibinfo{author}{\bibfnamefont{W.}~\bibnamefont{Losert}},
  \bibinfo{journal}{Europhys. Lett.} \textbf{\bibinfo{volume}{66}},
  \bibinfo{pages}{205} (\bibinfo{year}{2004}).

\bibitem[{\citenamefont{Alexander et~al.}(2004)\citenamefont{Alexander, Muzzio,
  and Shinbrot}}]{ale04}
\bibinfo{author}{\bibfnamefont{A.}~\bibnamefont{Alexander}},
  \bibinfo{author}{\bibfnamefont{F.~J.} \bibnamefont{Muzzio}},
  \bibnamefont{and} \bibinfo{author}{\bibfnamefont{T.}~\bibnamefont{Shinbrot}},
  \bibinfo{journal}{Gran. Matter} \textbf{\bibinfo{volume}{5}},
  \bibinfo{pages}{171} (\bibinfo{year}{2004}).

\bibitem[{\citenamefont{Charles et~al.}(2006)\citenamefont{Charles, Khan, and
  Morris}}]{cha06}
\bibinfo{author}{\bibfnamefont{C.~R.~J.} \bibnamefont{Charles}},
  \bibinfo{author}{\bibfnamefont{Z.~S.} \bibnamefont{Khan}}, \bibnamefont{and}
  \bibinfo{author}{\bibfnamefont{S.~W.} \bibnamefont{Morris}},
  \bibinfo{journal}{Gran. Matter} \textbf{\bibinfo{volume}{8}},
  \bibinfo{pages}{1} (\bibinfo{year}{2006}).

\bibitem[{\citenamefont{Aranson et~al.}(1999)\citenamefont{Aranson, Tsimring,
  and Vinokur}}]{ara99}
\bibinfo{author}{\bibfnamefont{I.~S.} \bibnamefont{Aranson}},
  \bibinfo{author}{\bibfnamefont{L.~S.} \bibnamefont{Tsimring}},
  \bibnamefont{and} \bibinfo{author}{\bibfnamefont{V.~M.}
  \bibnamefont{Vinokur}}, \bibinfo{journal}{Phys. Rev. E}
  \textbf{\bibinfo{volume}{60}}, \bibinfo{pages}{1975} (\bibinfo{year}{1999}).

\bibitem[{\citenamefont{Khan et~al.}(2004)\citenamefont{Khan, Tokaruk, and
  Morris}}]{kha04}
\bibinfo{author}{\bibfnamefont{Z.~S.} \bibnamefont{Khan}},
  \bibinfo{author}{\bibfnamefont{W.~A.} \bibnamefont{Tokaruk}},
  \bibnamefont{and} \bibinfo{author}{\bibfnamefont{S.~W.}
  \bibnamefont{Morris}}, \bibinfo{journal}{Europhys. Lett.}
  \textbf{\bibinfo{volume}{66}}, \bibinfo{pages}{212} (\bibinfo{year}{2004}).

\bibitem[{\citenamefont{Yanagita}(1999)}]{yan99}
\bibinfo{author}{\bibfnamefont{T.}~\bibnamefont{Yanagita}},
  \bibinfo{journal}{Phys. Rev. Lett.} \textbf{\bibinfo{volume}{82}},
  \bibinfo{pages}{3488} (\bibinfo{year}{1999}).

\bibitem[{\citenamefont{Shoichi}(1998)}]{sho98}
\bibinfo{author}{\bibfnamefont{S.}~\bibnamefont{Shoichi}},
  \bibinfo{journal}{Mod. Phys. Lett. B} \textbf{\bibinfo{volume}{12}},
  \bibinfo{pages}{115} (\bibinfo{year}{1998}).

\bibitem[{\citenamefont{Taberlet et~al.}(2004)\citenamefont{Taberlet, Losert,
  and Richard}}]{tab04}
\bibinfo{author}{\bibfnamefont{N.}~\bibnamefont{Taberlet}},
  \bibinfo{author}{\bibfnamefont{W.}~\bibnamefont{Losert}}, \bibnamefont{and}
  \bibinfo{author}{\bibfnamefont{P.}~\bibnamefont{Richard}},
  \bibinfo{journal}{Europhys. Lett.} \textbf{\bibinfo{volume}{68}},
  \bibinfo{pages}{522} (\bibinfo{year}{2004}).

\bibitem[{\citenamefont{Dury and Ristow}(1997)}]{dur97}
\bibinfo{author}{\bibfnamefont{C.~M.} \bibnamefont{Dury}} \bibnamefont{and}
  \bibinfo{author}{\bibfnamefont{G.~H.} \bibnamefont{Ristow}},
  \bibinfo{journal}{J. Phys. I (France)} \textbf{\bibinfo{volume}{7}},
  \bibinfo{pages}{737} (\bibinfo{year}{1997}).

\bibitem[{\citenamefont{Cundall and Strack}(1979)}]{cun79}
\bibinfo{author}{\bibfnamefont{P.~A.} \bibnamefont{Cundall}} \bibnamefont{and}
  \bibinfo{author}{\bibfnamefont{O.~D.~L.} \bibnamefont{Strack}},
  \bibinfo{journal}{G\'eotechnique} \textbf{\bibinfo{volume}{29}},
  \bibinfo{pages}{47} (\bibinfo{year}{1979}).

\bibitem[{\citenamefont{Haff and Werner}(1986)}]{haf86}
\bibinfo{author}{\bibfnamefont{P.~K.} \bibnamefont{Haff}} \bibnamefont{and}
  \bibinfo{author}{\bibfnamefont{B.~T.} \bibnamefont{Werner}},
  \bibinfo{journal}{Powder Tech.} \textbf{\bibinfo{volume}{48}},
  \bibinfo{pages}{239} (\bibinfo{year}{1986}).

\bibitem[{\citenamefont{Sch{\"a}fer et~al.}(1996)\citenamefont{Sch{\"a}fer,
  Dippel, and Wolf}}]{sch96}
\bibinfo{author}{\bibfnamefont{J.}~\bibnamefont{Sch{\"a}fer}},
  \bibinfo{author}{\bibfnamefont{S.}~\bibnamefont{Dippel}}, \bibnamefont{and}
  \bibinfo{author}{\bibfnamefont{D.~E.} \bibnamefont{Wolf}},
  \bibinfo{journal}{J. Phys. I (France)} \textbf{\bibinfo{volume}{6}},
  \bibinfo{pages}{5} (\bibinfo{year}{1996}).

\bibitem[{\citenamefont{Silbert et~al.}(2001)\citenamefont{Silbert, Ertas,
  Grest, Halsey, Levine, and Plimpton}}]{sil01}
\bibinfo{author}{\bibfnamefont{L.~E.} \bibnamefont{Silbert}},
  \bibinfo{author}{\bibfnamefont{D.}~\bibnamefont{Ertas}},
  \bibinfo{author}{\bibfnamefont{G.~S.} \bibnamefont{Grest}},
  \bibinfo{author}{\bibfnamefont{T.~C.} \bibnamefont{Halsey}},
  \bibinfo{author}{\bibfnamefont{D.}~\bibnamefont{Levine}}, \bibnamefont{and}
  \bibinfo{author}{\bibfnamefont{S.~J.} \bibnamefont{Plimpton}},
  \bibinfo{journal}{Phys. Rev. E} \textbf{\bibinfo{volume}{64}},
  \bibinfo{pages}{051302} (\bibinfo{year}{2001}).

\bibitem[{\citenamefont{Rapaport}(2004)}]{rap04}
\bibinfo{author}{\bibfnamefont{D.~C.} \bibnamefont{Rapaport}},
  \emph{\bibinfo{title}{The Art of Molecular Dynamics Simulation}}
  (\bibinfo{publisher}{Cambridge University Press},
  \bibinfo{address}{Cambridge}, \bibinfo{year}{2004}), \bibinfo{edition}{2nd}
  ed.

\bibitem[{\citenamefont{Ristow and Nakagawa}(1999)}]{rna99}
\bibinfo{author}{\bibfnamefont{G.~H.} \bibnamefont{Ristow}} \bibnamefont{and}
  \bibinfo{author}{\bibfnamefont{M.}~\bibnamefont{Nakagawa}},
  \bibinfo{journal}{Phys. Rev. E} \textbf{\bibinfo{volume}{59}},
  \bibinfo{pages}{2044} (\bibinfo{year}{1999}).

\bibitem[{\citenamefont{Levine}(1999)}]{lev99}
\bibinfo{author}{\bibfnamefont{D.}~\bibnamefont{Levine}},
  \bibinfo{journal}{Chaos} \textbf{\bibinfo{volume}{9}}, \bibinfo{pages}{573}
  (\bibinfo{year}{1999}).

\bibitem[{\citenamefont{Chandrasekhar}(1961)}]{cha61}
\bibinfo{author}{\bibfnamefont{S.}~\bibnamefont{Chandrasekhar}},
  \emph{\bibinfo{title}{Hydrodynamic and Hydromagnetic Stability}}
  (\bibinfo{publisher}{Oxford University Press}, \bibinfo{address}{Oxford},
  \bibinfo{year}{1961}).

\end{thebibliography}

\end{document}